\documentstyle[preprint,aps]{revtex}
\begin{document}
\draft
\title{Systematics of Gamow-Teller strengths in mid-$fp$-shell
nuclei}
\author{$\underline{\hbox{S. E. Koonin}}$ and K. Langanke}
\address{W. K. Kellogg Radiation Laboratory, 106-38\\
California Institute of Technology,
Pasadena, California 91125, USA}
\date{\today}
\maketitle
\begin{abstract}
We show that the presently available data on the Gamow-Teller (GT)
strength in
mid-$fp$-shell nuclei are proportional to the product of the numbers
of valence
protons and neutron holes in the full $fp$-shell. This observation
leads to
important insights into the mechanism for GT quenching and to a
simple
parametrization of the Gamow-Teller strengths important for electron
capture by
$fp$-shell nuclei in the early stage of supernovae.
\end{abstract}

\baselineskip=20pt plus1pt minus1pt

\narrowtext
Weak interactions play an important role in late stellar evolution.
In
particular, electron capture and nuclear beta decay are essential
during the
pre-supernova core collapse of a massive star, as these two processes
determine, in the early stage of the collapse, the electron-to-baryon
ratio. This quantity, in turn, influences both the infall dynamics
and
the mass of the final homologous core. Bethe {\it et al.}
\cite{Bethe} and
subsequently Fuller {\it et al.} \cite{Fuller} recognized that the
Gamow-Teller
(GT) resonance contributes significantly, and perhaps dominates, both
the
electron capture and beta decay rates in a collapsing star. GT
transition strengths in mid-$fp$-shell nuclei (atomic
numbers $Z \approx 22$--32) are therefore a key ingredient of
pre-supernova studies.

Small $Q$-values usually prevent laboratory studies of electron
capture and beta decay
from sampling the GT strength distribution over a wide range of
energies;
often, the GT resonance peak cannot be probed. However, the
intermediate-energy
$(n,p)$ charge-exchange reaction at forward angles is dominated by
the $\rm GT_+$
operator and so provides an alternative and energetically favorable
way of
determining the required strength function. In recent years the
CHARGEX
collaboration at TRIUMF has measured $(n,p)$ forward cross sections
for several
astrophysically important $fp$-shell nuclei. The total Gamow-Teller
strengths, $B({\rm GT}_+)$, extracted from these measurements are
summarized in
Table~\ref{Tab1}. These values typically sum the strength to
excitation
energies of 8~MeV and so span the energies of astrophysical interest;
it is generally believed that this low-energy range covers
most of the $\rm GT_+$ strength.

The GT transition is naively viewed as a one-body process in which a
proton is
changed into a neutron by the $\vec{\sigma} \tau_+$ operator. The
measured
${\rm GT}_+$ strengths are thus usually compared to an extreme single
particle
model \cite{Fuller}, in which the nuclear ground state is described
by
non-interacting nucleons occupying the lowest possible shell-model
orbitals.
The ${\rm GT}_+$ strength is then estimated as
\widetext
\begin{equation}
{B({\rm GT}_+)} =
\frac{1}{(2 j_i +1)}\vert\langle\vec{\sigma} \tau_+\rangle\vert^2 =
\sum_{i,f} \frac{n^p_i n^h_f}{(2 j_i+1)(2 j_f+1)}
\vert\langle i\vert\vec{\sigma} \tau_+\vert f\rangle\vert^2 \;,
\end{equation}
\narrowtext
where $n^p_i$ is the number of protons in orbital $i$ of the parent
ground state (spin $j_i$), $n^h_f$ is
the number of neutron holes in the final daughter orbital (with spin
$j_f$),
and the sum is over all proton orbitals in the initial state and
neutron orbitals in
the final states; transitions within the inert $A=40$ core are
blocked by the Pauli
principle. As the GT operator conserves orbital angular momentum, an
initial proton orbital can be connected to at most two different
final neutron
orbitals. The single particle matrix elements are given in
Ref.~\cite{Fuller}; they
are roughly equal for the $f_{7/2}$ and $p_{3/2}$ proton orbitals
that are important in the following discussion.

It is well known that the observed ${\rm GT}_+$ strength is
``quenched''; i.e.,
it is significantly smaller than the single particle estimate (1).
Aufderheide
{\it et al.} \cite{Aufderheide} justifiably criticized pre-supernova
studies that use
single-particle estimates in calculating electron-capture rates
\cite{Fuller},
However, it is also apparent from Table~\ref{Tab1} that the quenching
factor $q$ (ratio
of the single-particle estimate to experiment) varies significantly
from one
nucleus to another, so that Ref.~\cite{Aufderheide}'s suggestion of a
constant quenching, $q=2$, is also questionable. The purpose of this
Letter is
to point out a remarkable systematic behavior of the ${\rm GT}_+$
strengths for
mid-$fp$-shell nuclei that offers both insight into the quenching
mechanism and an improved estimate of astrophysical rates.

We begin by observing that the ${\rm GT}_+$ strength should be
proportional to
the number of valence protons in the $fp$-shell, $Z_{\rm val}=Z-20$.
For two
reasons, the ${\rm GT}_+$ strength also depends upon the number of
valence
neutrons, $N_{\rm val}=N-20$. First, the valence neutrons block
possible
transitions, as is accounted for in Eq.~(1). Second, neutron-proton
correlations (not included in the single-particle estimate) have been
identified as a major source of quenching \cite{Vogel,Muto}.
Motivated by these
considerations, we divide the experimental $B({\rm GT}_+)$ values in
Table~\ref{Tab1} by
the number of valence protons and plot them as a function of $N_{\rm
val}$ in Fig.~\ref{fig1}. (We omit the lightest nucleus, ${}^{45}$Sc,
for reasons discussed below.) As expected, the values decrease with
increasing $N_{\rm val}$ and do so roughly linearly. We are therefore
led to parametrize the measured total ${\rm GT}_+$ strengths for
mid-shell-$fp$-nuclei as
\begin{equation}
B({\rm GT}_+)=a\cdot Z_{\rm val}\cdot (b-N_{\rm val})\;.
\end{equation}

A closer inspection of the data (see Fig.~\ref{fig1}) reveals that
$B({\rm GT}_+)/Z_{\rm val}$ for the odd-$Z$ nuclei ${}^{51}$V,
${}^{55}$Mn, and ${}^{59}$Co is systematically lower than the values
for the neighboring even-$Z$ nuclei, so that the measurements
strongly suggest an odd-even dependence. However, we believe that
this behavior is caused mainly by kinematics. The $(n,p)$ experiments
can reliably determine the $B({\rm GT}_+)$ strength only up to
daughter excitation energies of about $E_x=8$~MeV. However, the GT
resonance appears in the $(n,p)$ spectra at systematically higher
excitation energies for odd-$Z$ targets than for even-$Z$ targets
\cite{Auf,Jack}. Experiments with odd-$Z$ targets will therefore
``miss'' a relatively larger fraction of the total strength with
$E_x>8$~MeV.

Two additional points support our interpretation of this apparent
odd-even effect. First, the data for ${}^{51}$V and ${}^{58}$Co show
additional $B({\rm GT}_+)$ strength at higher excitation energies
between $E_x=8$ and 12~MeV \cite{Jackson}. Aufderheide {\it et al.}
\cite{Auf} have analyzed the data on ${}^{51}$V, ${}^{54}$Fe, and
${}^{58}$Co, and cite $B({\rm GT}_+)$ strengths extending to higher
excitation energies ($E_x\approx12$~MeV) for ${}^{51}$V, ${}^{54}$Fe,
and ${}^{59}$Co. While the ${}^{54}$Fe result coincides with the
value summed to $E_x=8$~MeV given in Ref.~\cite{Fe}, their values for
${}^{51}$V and ${}^{59}$Co are both noticeably larger than those of
Ref.~\cite{Alford} (which only give the strengths up to $E_x=8$~MeV,
see Table~\ref{Tab1}) and both agree well with the linear systematics
deduced for the even-$Z$ nuclei (Fig.~\ref{fig1} and below). Second,
we discuss below recent shell model calculations that agree well with
the experimental $B({\rm GT}_+)$ values for even-$Z$ nuclei and do
not exhibit an odd-even dependence \cite{Koonin}. For the only
odd-$Z$ nucleus studied by these methods to date, ${}^{55}$Mn, the
calculated total $B({\rm GT}_+)$ per valence proton, $B({\rm
GT}_+)/Z_{\rm val}=0.44$, is in accord with the results for the
neighboring even-$Z$ nuclei ${}^{54}$Cr and ${}^{56}$Fe. Moreover,
the shell model study finds the centroid of the $B({\rm GT}_+)$
strength in ${}^{55}$Mn at higher daughter excitation energies
($E_x\approx2.4\pm1.6$~MeV) than in ${}^{56}$Fe
($E_x\approx-0.4\pm0.2$~MeV), in agreement with both the data and our
interpretation.

We therefore suggest that the {\it total} $B({\rm GT}_+)$ strengths
of mid-$fp$-shell nuclei follow the simple parametrization as given
in Eq.~(2), while the odd-even behavior of the data reflects the
experimental excitation energy cut-off at around 8~MeV. As can be
seen from the line in Fig.~\ref{fig1}, the presently available
$B({\rm GT}_+)$ data are well-fitted by Eq.~(2) if the ``total''
$B({\rm GT}_+)$ values for ${}^{51}$V and ${}^{59}$Co given in
Ref.~\cite{Auf} are used. We then find the best-fit values
$a=(4.55\pm0.25)\times10^{-2}$ and $b=19.54\pm0.32$ with $\chi^2=1.0$
per degree of freedom. A similarly good fit is obtained if $b$ is
constrained to be 20, where we find $a=(4.29\pm0.15)\times10^{-2}$
with $\chi^2=1.1$ per degree of freedom. The parameter $a$ can be
interpreted as an average matrix element. Note that fitting the same
$B({\rm GT}_+)$ data to Eq.~(1) multiplied by a constant quenching
factor results in $q\approx3.7$ with $\gamma^2=2.4$ per degree of
freedom, which is a noticeably worse fit to the data than provided by
Eq.~(2).

We expect our parametrization (2) to be valid for nuclei between
${}^{48}$Ti and
${}^{70}$Ge; i.e., for $Z$ between 22 and 32 and $N$ between
approximately 26 and
38. Most of the nuclei whose electron capture and beta-decay rates
govern
the late stage of stellar evolution are within these ranges. However,
it should
be noted that our parametrization has been derived solely from
experiments with
even-$N$ targets, so that its validity for odd-$N$ nuclei remains to
be
verified. The parametrization is clearly not valid for nuclei with
neutron
numbers in excess of 40 (which require the inclusion of the $g_{9/2}$
orbital, but will have very small $\rm GT_+$ strengths for
$Z=22$--32) nor for
nuclei with only a few valence nucleons, where we do not expect the
proton-neutron
correlations to be fully developed. For example, for $N_{\rm val}=0$
$B({\rm GT}_+)/Z_{\rm val}=3$ and there should be no quenching. For
${}^{45}$Sc
($Z_{\rm val}=1$) the experimentally observed quenching ($q=1.1$) is
also much less than predicted by Eq.~(2) ($q=3.4$).

Several interesting conclusions follow from the validity of the
parametrization (2).
\begin{enumerate}
\item
The ${\rm GT}_+$ strength is proportional to (20-$N_{\rm val}$), the
number of
neutron holes in the {\it full} $fp$-shell. This indicates that the
strength is
determined by the {\it total} number of holes, rather than by the
numbers of
holes in individual subshells (as is assumed in Eq.~(1)), and
suggests that
proton-neutron correlations are a significant determinant of the
${\rm GT}_+$
strength. ${\rm GT}_+$ strengths of mid-$fp$-shell nuclei apparently
behave as
though there was only one large shell in which all sub-shell
structures have
been diluted.

\item
The $N_{\rm val}$-dependence of our parametrization suggests that
correlations
introduced by higher shells (e.g., the $g_{9/2}$ orbital) do not
significantly
change the $B({\rm GT}_+)$ values for the range of nuclei covered by
the
empirical formula (2).

\item
The apparent sensitivity of $B({\rm GT}_+)$ to the number of neutron
holes in the
{\it full} $fp$-shell suggests that shell model calculations
attempting to
reproduce the GT strength must include the full $fp$-shell model
space. This
observation is in accord with recent complete $0\hbar\omega$
$fp$-shell Monte Carlo
calculations \cite{Koonin}, which show quenching significantly
greater than
that calculated in restricted ($2p$-$2h$) model spaces due to
proton-neutron
correlations \cite{Dean}. In fact, these calculations, employing the
Brown-Richter interaction \cite{Brown}, yield $B({\rm GT}_+)$ values
for
${}^{54}$Cr, ${}^{54}$Fe, ${}^{55}$Mn, and ${}^{56}$Fe that agree
well with
experiment and with the empirical parametrization (see
Fig.~\ref{fig1}). Only the shell
model value for ${}^{58}$Ni is significantly larger than experiment,
a result
that is quite sensitive to the Hamiltonian assumed. A recent complete
$0\hbar\omega$ direct
diagonalization of ${}^{48}$Ti yields a $B({\rm GT}_+)$ value of
$1.26=0.045Z_{\rm val}(20-N_{\rm val})$ \cite{Zuker}, again in
agreement with the data
and Eq.~(2).

\item
Our attempt at a similar parametrization for $sd$-shell nuclei
failed,
indicating that the GT strengths in these nuclei are more sensitive
to the sub-shell structure.

\end{enumerate}

In summary, we have observed that presently available $(n,p)$ data
for mid-$fp$-shell
nuclei show that the total ${\rm GT}_+$ strength is proportional to
the
product of the number of valence protons and the number of neutron
holes in the
full $fp$-shell. This observation suggests the importance of
neutron-proton
correlations throughout all of the $fp$ orbitals in reproducing the
observed
quenching. It also leads to a simple empirical parametrization of the
${\rm
GT}_+$ strength valid for all nuclei between ${}^{48}$Ti and
${}^{70}$Ge. As the
electron capture and beta-decay rates for nuclei in this range
determine the
early stages of a supernova collapse, our parametrization will help
to
reduce the uncertainties in pre-supernova studies. We also note that
the odd-even dependence of the excitation energy of the GT resonance
should be quite important in astrophysical applications, but has
apparently been neglected in supernova studies to date.

\acknowledgements

We thank the CHARGEX collaboration for providing us with their
$(n,p)$ data prior
to publication and are grateful to Petr Vogel for helpful
discussions. This work was supported in part by the National Science
Foundation, Grants No. PHY91-15574 and PHY90-13248.

\begin{figure}
\caption{Plot of the experimental Gamow-Teller strength, $B({\rm
GT}_+)$, per valence
proton
as a function of the number of valence neutrons. Open symbols with
error bars denote data summed to $E_x\approx8$~MeV, while the full
symbols represent the Gamow-Teller strengths summed to
$E_x\approx12$~MeV for the odd-$Z$ nuclei ${}^{51}$V and ${}^{59}$Co
\protect\cite{Auf}. Symbols without error bars
denote recent
full-space shell model results, as discussed in the text.
The line shows the best fit to Eq.~(2) when $b$ is constrained
to be 20 and the adjusted data for ${}^{51}$V and ${}^{59}$Co are
used.}
\label{fig1}
\end{figure}

\widetext
\begin{table}
\caption{Comparison of measured Gamow-Teller strengths for
$E_x<8$~MeV extracted
from $(n,p)$ data on midshell $fp$-nuclei with the prediction
of the single particle model (1). The quenching
factor $q$ is defined as the ratio of these two quantities. For
${}^{51}$V and ${}^{59}$Co the ``total'' Gamow-Teller strength for
$E_x<12$~MeV as given in Ref.~\protect\cite{Auf} is also listed.}
\begin{tabular}{ccd@{}l@{}ldd@{}l}
Target & $(Z,N)$ &
\multicolumn{3}{c}{$B({\rm GT}_+)$} &
\multicolumn{1}{c}{Single particle estimate} &
\multicolumn{2}{c}{$q$}\\
&&&&&Eq. (1)\\
\tableline
${}^{45}$Sc & (21,24) &
2.1 & & \protect\cite{Sc} &
2.36 &
1.1\\
${}^{48}$Ti & (22,26) &
1.31 & $\pm 0.2$ & \protect\cite{Ti} &
4.07 &
3.1 & $\pm 0.5$ \\
${}^{51}$V & (23,28) &
1.2 & $\pm 0.15$ & \protect\cite{Alford} &
5.14 &
4.3 & $\pm 0.4$ \\
&&
1.48 & $\pm0.15$ & \protect\cite{Auf}&
&
3.5 & $\pm0.4$ \\
${}^{54}$Fe & (26,28) &
3.1 & $\pm 0.6$ & \protect\cite{Fe} &
10.29 &
3.3 & $\pm 0.6$ \\
${}^{55}$Mn & (25,30) &
1.72 & $\pm 0.2$ & \protect\cite{Jack} &
8.57 &
5.0 & $\pm 0.5$ \\
${}^{56}$Fe & (26,30) &
2.85 & $\pm 0.3$ & \protect\cite{Jack} &
10.29 &
3.6 & $\pm 0.4$ \\
${}^{58}$Ni & (28,30) &
3.76 & $\pm 0.40$ & \protect\cite{Jack} &
13.71 &
3.6 & $\pm 0.2$ \\
${}^{59}$Co & (27,32) &
1.9 & $\pm 0.2$ & \protect\cite{Alford} &
12.02 &
6.3 & $\pm 0.4$ \\
&&
2.39 & $\pm0.25$ & \protect\cite{Auf} &
&
5.0 & $\pm 0.5$ \\
${}^{70}$Ge & (32,38) &
0.84 & $\pm 0.13$ & \cite{Ge} &
2.67 &
3.2 & $\pm 0.5$ \\
\end{tabular}
\label{Tab1}
\end{table}
\narrowtext

\end{document}